\documentstyle[12pt,twoside,fleqn,espcrc1,epsfig]{article}

\newcommand{\AmS}{{\protect\the\textfont2
  A\kern-.1667em\lower.5ex\hbox{M}\kern-.125emS}}

\title{Covariant Description of Composite Meson Systems 
and Chiral Symmetry}
\author{Shin Ishida,\address{Coll. Sci. and Tech. Nihon U., 
Kanda-Surugadai, Chiyoda, Tokyo 101-0062, Japan}
Muneyuki Ishida\address{Department of Physics, 
Tokyo Institute of Technology,Tokyo 152-8551, Japan} 
and Tomohito Maeda$^{\rm a}$  }
       
\begin{document}

\maketitle

\begin{abstract}
Assuming the spin-independence for confining force, we give a covariant
quark 
representation of general composite meson systems with definite 
Lorentz transformation properties. For benefit of this representation
we are able to deduce automatically the transformation rules of composite
mesons for general symmetry operations from those of constituent (exciton)
quarks. Applying this we investigate especially physical implication 
of chiral symmetry for the meson systems,
and point out a possibility of existence of new meson multiplets.
\end{abstract}

\section{Introduction}
There are the two contrasting view points of composite quark-antiquark
mesons: The one is non-relativistic, based on the approximate symmetry of
$LS$-coupling in NRQM; while the other is relativistic, based on the 
dynamically broken chiral symmetry in the NJL model. 
The $\pi$-meson (or $\pi$-nonet) is now widely believed to have a
dual nature of non-relativistic particle with $(L,S)=(0,0)$
and also of relativistic particle as a Nambu-Goldston boson with
$J^P=0^-$ in the case of spontaneous breaking of chiral symmetry. However,
no attempts to unify the above two view points have been yet proposed.
On the other hand we have developed the 
covariant oscillator quark model (COQM)\cite{rf1} 
for many years as a covariant extension of NRQM, which is based on 
the boosted $LS$-coupling scheme. The meson wave functions (WF)
in COQM are tensors in the $\tilde U(4)\bigotimes O(3,1)$ space
and reduce at the rest frame to those in the 
$SU(2)_{\rm spin}\bigotimes O(3)_{\rm orbit}$ space in NRQM.
The COQM has been successful especially in treating the
$Q\bar Q$ meson system and the 
($q,\bar Q$) or ($Q,\bar q$) meson system, leading, respectively, 
to a satisfactory understanding of radiative transitions 
and to the same
weak form factor relations as in HQET.
However, in COQM no 
consideration on chiral symmetry has been given and it is not able to 
explain the dual nature of $\pi$ meson.

The purpose of the present talk is to get rid of this defect in COQM and is to give a unified view point of the two contrasting view points
of the composite meson systems.

\section{Covariant Framework for Describing Composite Mesons}
For meson WF described by $\Phi_A{}^B(x_1,x_2)$ ($x_1,x_2$ denoting the space-time coordinate and 
$A=(\alpha ,a)(B=(\beta ,b))$ denoting the Dirac spinor and 
flavor indices  of constituent quark (anti-quark)) 
we set up the bilocal Yukawa equation\cite{rf2}
\begin{eqnarray}
[\partial^2/\partial X_\mu^2
 &-& {\cal M}^2(x_\mu ,\partial/\partial x_\mu )]
\Phi_A{}^B(X,x)=0
\label{eq1}
\end{eqnarray}
($X(x)$ denoting the center of mass (CM) (relative) coordinate of meson),
where the  ${\cal M}^2$ is squared mass operator including
only a central, (Dirac) spinor\\
-independent confining potential.
The WF is separated into the plane wave describing 
CM motion and the internal WF as 
\begin{eqnarray}
\Phi_A{}^B(x_1,x_2)=\sum_{{\bf P}_n,n}(e^{iP_nX}\Psi_{n,A}{}^{(+)B}(x,P_n)
+e^{-iP_nX}\Psi_{n,A}{}^{(-)B}(x,P_n)),
\label{eq2}
\end{eqnarray}
where $P_{n,\mu}^2=-M_n^2,\ P_{n,0}=\sqrt{M_n^2+{\bf P}_n^2}$;
$(\pm )$ represents the positive (negative) frequency part;
and $n$ does a freedom of excitation. We have the following field 
theoretical expression in mind for the WF and its Pauli-conjugate,
$\bar\Phi_B{}^A(x_2,x_1)\equiv [\gamma_4\Phi (x_1,x_2)^\dagger\gamma_4
]_B{}^A$
as
\begin{eqnarray}
\Phi_A{}^B(x_1,x_2) &=& \sum_n[\langle 0|\psi_A(x_1)\bar\psi^B(x_2)|M_n\rangle
+\langle M_n|\psi_A(x_1)\bar\psi^B(x_2)|0\rangle ],\nonumber\\
\bar\Phi_B{}^A(x_2,x_1) &=& \sum_n[
\langle M_n|\psi_B(x_2)\bar\psi^A(x_1)|0\rangle
+\langle 0|\psi_B(x_2)\bar\psi^A(x_1)|M_n\rangle ],
\label{eq3}
\end{eqnarray}
where $\psi_A(\bar\psi^B)$ denotes the quark field (its Pauli-conjugate)
and $|M_n\rangle$ does the composite meson state.
The internal WF is, concerning the spin-dependence, expanded in terms of 
a complete set $\{ W^{(i)} \}$ of free bi-Dirac spinors of 
quarks and anti-quarks; and 
the Fierz-component meson WF is expressed as
\begin{eqnarray}
\Psi_A^{(\pm )B}(x,P_n)=\sum_{(i)}W_\alpha^{(i)(\pm )\beta}(P_n)
M_a^{(i)(\pm )b}(x,P_n),\ \ 
M^{(i)(\pm )}(x,P_n)=\epsilon^{(i)}\langle\bar W^{(i)(\mp )}\Psi^{(\pm)}
\rangle ,
\label{eq4}
\end{eqnarray}
where $\langle A\rangle$ means trace of $A$ and ortho-normal
relations $\langle\bar W^{(i)(\mp )}W^{(j)(\pm )} \rangle 
=\epsilon^{(i)}\tilde\delta_{ij}$
($\epsilon^{(i)}$ and $\tilde\delta_{ij}$ denote,respectively, the sign
and pseudo-Kronecker symbols) is used.

The meson WF satisfies, as is seen from Eq.(\ref{eq3}),
the self-conjugate condition, leading to the (conventional) 
crossing rule (or substituion law)
for the (composite) meson WF, as
\begin{eqnarray}
\bar\Phi_A{}^B(x_1,x_2)=\Phi_A{}^B(x_1,x_2),\ \ 
\bar M_b^{(i)(\pm )a}(-x,P_n)=M_a^{(i)(\mp )b}(x,P_n)^\dagger
=M_b^{(i)(\pm )a}(x,P_n).
\label{eq6}
\end{eqnarray}

\section{Complete Set of Spin Wave Function
and Heavy-quark Meson Systems}
We set up the conventional ``free'' Dirac spinors with four-momentum of 
composite meson itself $P=P_M$,
$D_{q,\alpha}(P)\equiv (u_{q,\alpha}^{(r)}(P),v_{q,\alpha}^{(r)}(P)
(r=1,2))$ for quarks and  
$\bar D_{\bar q}{}^\beta (P)\equiv  (\bar v_{\bar q}^{(s)\beta}(P),
\bar u_{\bar q}^{(s)\beta}(P)(s=1,2))$ for anti-quarks.
It is to be noted that all four spinors are necessary for both quarks 
and anti-quarks inside of mesons. Then the complete set of 
bi-Dirac spinors is 
given by\footnote{
In the following we give only the expressions of $(+)$-frequency parts,
and consider only the ground states of composite system, disregarding the 
relative coordinates.
}
\begin{eqnarray}
\{ W^{(i)(+)}(P) \} &:& 
U(P)=u_q^{(r)}(p_1)\bar v_{\bar q}^{(s)}(p_2)|_{p_{i,\mu}=\kappa_iP_\mu}
=u_+^{(r)}({\bf P})\bar v_+^{(s)}({\bf P}),\nonumber\\
 &\ & 
C(P)=u_q^{(r)}(p_1)\bar u_{\bar q}^{(s)}(p_2)|_{p_{i,\mu}=\kappa_iP_\mu}
=u_+^{(r)}({\bf P})\bar v_-^{(s)}(-{\bf P}),\nonumber\\
 &\ & 
D(P)=v_q^{(r)}(p_1)\bar v_{\bar q}^{(s)}(p_2)|_{p_{i,\mu}=\kappa_iP_\mu}
=u_-^{(r)}(-{\bf P})\bar v_+^{(s)}({\bf P}),\nonumber\\
 &\ & 
V(P)=v_q^{(r)}(p_1)\bar u_{\bar q}^{(s)}(p_2)|_{p_{i,\mu}=\kappa_iP_\mu}
=u_-^{(r)}(-{\bf P})\bar v_-^{(s)}(-{\bf P}),
\label{eq9}
\end{eqnarray}
where
$u_+({\bf P})(\bar v_+({\bf P}))$ and 
$u_-(-{\bf P})(\bar v_-(-{\bf P}))$ denote the Dirac spinors with 
positive energy and momentum ${\bf P}$ and with negative energy 
and momentum $-{\bf P}$, respectively, describing quarks 
(anti-quarks). In Eq.(\ref{eq9}) 
we have defined technically the momenta of
constituent quarks (to be called quark-exciton)\cite{rf3} as 
\begin{eqnarray}
p_{i,\mu} \equiv  \kappa_iP_\mu ,\ p_{i,\mu}^2=-m_i^2;\ P_\mu^2=-M^2,
M=m_1+m_2\ (\kappa_{1,2}\equiv m_{1,2}/(m_1+m_2)).
\ \ \ \ \ \ \ \ \ \ \ \ 
\label{eq10}
\end{eqnarray}
The respective members in Eq.(\ref{eq9})
satisfy a couple of the corresponding free Dirac equations in momentum 
space (which are equivalent to the (conventional or 
new-type of) Bargman-Wigner Equations) and are expressed 
in terms of their irreducible composite meson WF 
as follows:
\begin{eqnarray}
{\rm (Non\ relat.\  comp.)} & & 
  U_A{}^B(P) = \frac{1}{2\sqrt{2}}
   [(i\gamma_5P_{s,a}^{(NR)b}(P)+i\gamma_\mu V_{\mu ,a}^{(NR)b}(P))
    (1+\frac{iP\gamma}{M}) ]_\alpha{}^\beta ,\nonumber \\
{\rm (Semi\ relat.\ comp.)} & & 
  C_A{}^B(P) = \frac{1}{2\sqrt{2}}
   [(S_{a}^{(\bar q) b}(P)+i\gamma_5\gamma_\mu A_{\mu ,a}^{(\bar q)b}(P))
    (1-\frac{iP\gamma}{M}) ]_\alpha{}^\beta ,\nonumber\\
  & & D_A{}^B(P) = \frac{1}{2\sqrt{2}}
   [(S_{a}^{(q) b}(P)+i\gamma_5\gamma_\mu A_{\mu ,a}^{(q)b}(P))
    (1+\frac{iP\gamma}{M}) ]_\alpha{}^\beta ,\nonumber\\
{\rm (Extrly.relat.\ comp.)} & & 
  V_A{}^B(P) = \frac{1}{2\sqrt{2}}
   [(i\gamma_5P_{s,a}^{(ER)b}(P)+i\gamma_\mu V_{\mu ,a}^{(ER)b}(P))
    (1-\frac{iP\gamma}{M}) ]_\alpha{}^\beta ,
\label{eq11}
\end{eqnarray}
where  
all vector and axial-vector mesons satisfy the Lorentz 
conditions, $P_\mu V_\mu (P)=P_\mu A_\mu (P)=0$.
Here it is to be noted that, in each type of the above members, 
the number of freedom counted both in the quark representation and 
in the meson representation is equal,
as it should be ($2\times 2=4$ and $1+3=4$, respectively).
Also it may be amusing to note that each constituent 
in all the above members is in 
``parton-like motion,''
having the same 3-dimentional velocity as that of total mesons.
(For example, in $V(P),\ {\bf v}_{1,2}=
\frac{{\bf p}^{(1,2)}}{p_0^{(1,2)}}=
\frac{-\kappa_{1,2}{\bf P}_M}{-\kappa_{1,2}P_{M,0}}=
{\bf v}_M$.)

In the heavy quarkonium ($Q\bar Q$) system both quarks and anti-quarks
are possible to do, since $m_Q > \Lambda_{\rm conf}$, 
only non-relativistic motions with positive energy.
Accordingly the bi-spinor $U$ is considered to be applied to $Q\bar Q$
 system as a covariant spin WF.
In the heavy-light quark meson $Q\bar q$($q\bar Q$) system
the anti-quarks(quarks) make, since   $m_q \ll \Lambda_{\rm conf}$, 
relativistic motions both with positive and negative energies.
Accordingly both the bi-spinors $U$ and $C$ ($U$ and $D$)
are to be applied to the  $Q\bar q$($q\bar Q$) system, and in this system
there is a possibility of 
existence of new composite scalar and axial-vector mesons(see
Eq.(\ref{eq11})).
In the light quark meson $q\bar q$-system both the quarks and anti-quarks
make, since $m_q\ll\Lambda_{\rm conf}$, relativistic motions both
with positive and negative energies.
Accordingly the (linear combinations to be specified shortly of)
bispinors $U$ and $V$ are applied to the   $q\bar q$-system, and
in this system there is a possibility of existence of an extra(, in 
addition, to a normal)  set of 
composite pseudo-scalar and vector mesons.

\section{Light-Quark Meson Systems and Chiral Symmetry}
\hspace*{-0.8cm}{\bf A)[Charge conjugation]} properties of the bi-spinors 
and, correspondingly, of the composite mesons are derived 
from those of quarks as follows:
\begin{eqnarray}
    \Psi_A^{(+)B} (P,x) 
 &(&\approx \langle 0|\psi_A(x_1)\bar\psi^B(x_2)|M\rangle )
   \rightarrow \Psi_A^{c,(+)B} (P,x)  
 (\approx \langle 0|\psi_A(x_1)\bar\psi^B(x_2)|M^c\rangle )\nonumber\\
   &=& (C^{-1})_{AA'} {}^t\Psi^{(+)}(P,-x)^{A'}{}_{B'}C^{B'B},\ \ 
            (CC^\dagger =1,\ C\gamma_\mu C^{-1}=-^t\gamma_\mu  
             );    \nonumber \\
    P_{s,a}^{(NR)b} &\leftrightarrow& P_{s,b}^{(NR)a},\   
      V_{\mu ,a}^{(NR)b} \leftrightarrow -V_{\mu ,b}^{(NR)a};
      S_{a}^{(R\bar q)b} \leftrightarrow S_{b}^{(Rq)a},\ 
   A_{\mu ,a}^{(R\bar q)b} \leftrightarrow A_{\mu ,b}^{(Rq)a};\nonumber\\
   P_{s,a}^{(ER)b} &\leftrightarrow& P_{s,b}^{(ER)a},\ 
      V_{\mu ,a}^{(ER)b} \leftrightarrow -V_{\mu ,b}^{(ER)a}.
\label{eq12}
\end{eqnarray}
{\bf B)[Chiral transformation]} properties of composite mesons are 
also derived straightforwardly from those of the bi-spinors 
\begin{eqnarray}
\psi_A{}^B(P,x) &\rightarrow & 
   [e^{i\alpha^i\frac{\lambda^i}{2}\gamma_5}\psi (P,x)
    e^{i\alpha^i\frac{\lambda^i}{2}\gamma_5}]_A{}^B.
\label{eq18}
\end{eqnarray}
{\bf C)[Light-quark meson system-``chiral SU(6) multiplet'']}\\
The quark representation applying to the 
light-quark mesons is obtained by the linear transformation of 
the bi-Dirac spinors given in \S 3 as follows:
\begin{eqnarray}
U_{P_s,\alpha}^{(N,E)\beta} &\equiv & 1/\sqrt{2}\ 
   (U_{P_s}\pm V_{P_s})_\alpha{}^\beta 
   = i/2\ [\gamma_5(1,iv\gamma )]_\alpha{}^\beta ;\ \ \ 
        P_s^{(N,E)}; C=(+,+) \nonumber\\
C_{S,\alpha}^{(N,E)\beta} &\equiv & (1,i)/\sqrt{2}\ 
   (D_S\pm C_S)_\alpha{}^\beta 
   = 1/2\ [(1,-v\gamma )]_\alpha{}^\beta ;\ \ \ 
        S^{(N,E)}; C=(+,-) \nonumber\\
U_{V,\alpha}^{(N,E)\beta} &\equiv & 1/\sqrt{2}\ 
   (U_{V}\pm V_{V})_\alpha{}^\beta 
   = i/2\ [(\tilde\gamma_\mu ,-\sigma_{\mu\nu}v_\nu )]_\alpha
    {}^\beta ;\ \ \   V^{(N,E)}; C=(-,-) \nonumber\\
C_{A,\alpha}^{(N,E)\beta} &\equiv & (1,i)/\sqrt{2}\ 
   (D_A\pm C_A)_\alpha{}^\beta 
   = i/2\ [\gamma_5(\tilde\gamma_\mu ,-i\sigma_{\mu\nu}v_\nu )
   ]_\alpha{}^\beta ;\ \ \   A^{(N,E)}; C=(+,-)
\label{eq20}
\end{eqnarray}
($v_\mu\equiv P_\mu /M,\ \tilde\gamma_\mu v_\mu\equiv 0$; and
 $U_{P_s}$ denotes the coefficient bi-spinors of $P_s$ and so on ),
where we have given also the charge-conjugation parity of the 
corresponding (hidden flavor) composite mesons. 
The chiral transformation properties
 of the new bi-spinors (for $U_L(1)\times U_R(1)$) are easily seen to 
be similar as the conventional ones as
\begin{eqnarray}
 &\  & 1 \leftrightarrow  i\gamma_5,\ \ 
-v\gamma\leftrightarrow -\gamma_5v\gamma ,\ \ 
 i\gamma_5\tilde\gamma_\mu 
  \leftrightarrow  i\tilde\gamma_\mu ,\ \ 
-i\sigma_{\mu\nu}v_\nu\leftrightarrow -\gamma_5
\sigma_{\mu\nu}v_\nu .
\label{eq21}
\end{eqnarray}
{\bf D)[``Local chiral SU(6) field'']}\\
Extending our considerations to include the $(-)$-frequency part,
we are led to a unified expression of what to be called,
Local Chiral SU(6) field, as
\begin{eqnarray}
\Psi_A{}^B(X) &=& \Psi_A^{(N)B}(X) + \Psi_A^{(E)B}(X) \nonumber\\
\Psi_A^{(N)B} &=& 1/2\ [
 i\gamma_{5\alpha}{}^\beta P_{s,a}^{(N)b}
 +i\tilde\gamma_{\mu ,\alpha}{}^\beta V_{\mu ,a}^{(N)b}
 +1_alpha{}^\beta S_{a}^{(N)b}
 +(i\gamma_5\tilde\gamma_\mu )_\alpha{}^\beta A_{\mu ,a}^{(N)b}]
 \nonumber\\
\Psi_A^{(E)B} &=& 1/2\ [
(i\gamma_5\gamma_\mu )_\alpha{}^\beta M^{-1}\partial_\mu 
  P_{s,a}^{(E)b}
  +(\sigma_{\mu\nu})_{\alpha}{}^\beta  M^{-1}\partial_\mu
  V_{\nu ,a}^{(E)b} \nonumber \\
 & &  +  
i\gamma_{\mu ,\alpha}{}^\beta M^{-1}\partial_\mu S_{a}^{(E)b}
    +(i\gamma_5\sigma_{\mu\nu})_{\alpha}{}^\beta  M^{-1}\partial_\mu
  A_{\nu ,a}^{(E)b}].
\label{eq22}
\end{eqnarray}

\section{Concluding remarks}
In this talk we have pointed out a possibility of existence of the new 
composite meson multiplets, when both or either constituent 
quark-excitons have ``negative'' energies and momenta. Here we should like
to point out that the chiral symmetry of QCD guarantees 
this to be the case, and there exist some candidates (to be discussed
elsewhere) for these new multiplets.


\begin{thebibliography}{9}
\bibitem{rf1}S. Ishida, M. Y. Ishida and M. Oda, 
   {\it Prog. Theor. Phys.} {\bf 93}, 939 (1995). 
\bibitem{rf2}H. Yukawa,
   {\it Phys. Rev.} {\bf 91}, 415 and 416 (1953). 
\bibitem{rf3}S. Ishida,
    {\it Prog. Theor. Phys.} {\bf 46}, 1570 and 1950 (1971). 
\end{thebibliography}
\end{document}